\documentclass[twocolumn,aps,showpacs,groupedaddress,floatfix,prl]{revtex4-1}
\usepackage{amsmath}
\usepackage{graphicx}
\usepackage{dcolumn}
\usepackage{bm}
\usepackage{verbatim}

\begin{document}
\setlength{\parskip}{0mm}
\title{Microscopic Origin of Shear Relaxation in Strongly Coupled Yukawa Liquids}
\author{Ashwin J.}
\email{ashwin@ipr.res.in}
\author{Abhijit Sen}
\affiliation{Institute for Plasma Research, Bhat, Gandhinagar-382428, India.}
\date{\today}
\begin{abstract}
We report accurate molecular dynamics calculations of the shear stress relaxation in a two-dimensional strongly coupled Yukawa liquid over a wide
range of the Coulomb coupling strength $\Gamma$ and the Debye screening parameter $\kappa$. Our data on the relaxation times of the ideal- , excess-
and total shear stress auto-correlation ($\tau^{id}_M, \tau^{ex}_M, \tau_M$ respectively) along with the lifetime of local atomic connectivity
$\tau_{LC}$ leads us to the following important observation. Below a certain crossover $\Gamma_c(\kappa)$, $\tau_{LC} \rightarrow \tau^{ex}_M$,
directly implying that here $\tau_{LC}$ is the microscopic origin of the relaxation of excess shear stress unlike the case for ordinary liquids where it
is the origin of the relaxation of the total shear stress. At $\Gamma >> \Gamma_c(\kappa)$ i.e. in the potential energy dominated regime,
$\tau^{ex}_M\rightarrow \tau_M$ meaning that $\tau^{ex}_M$ can fully account for the elastic or ``solid like'' behavior. 

\end{abstract} 
\pacs{52.27.Lw,51.35.+a,52.65.Yy}
\keywords{}
\maketitle
The Yukawa liquid is routinely used to model a wide variety of strongly coupled systems such as laboratory and astrophysical dusty plasma systems,
charged colloids and liquid metals \cite{RevModPhys.81.1353,Fortov20051}. The particle interaction potential used in this liquid has the form $\phi(r)
= \frac{Q^2}{4\pi\epsilon_0 r}e^{-\frac{r}{\lambda_D}}$. Here $Q$ and $\lambda_D$ refer to the particle charge and Debye shielding distance
respectively. The thermodynamic state point of the Yukawa liquid is completely characterized by two dimensionless quantities namely, the screening
parameter $\kappa = \frac{a}{\lambda_D}$ and the coupling strength $\Gamma = \frac{Q^2}{4\pi\epsilon_0 a k_B T}$. Here $a$ is the Wigner-Seitz radius
such that $\pi a^2 n = 1$ with $n$ being the areal number density. In a strongly coupled Yukawa liquid (SCYL), the average Coulomb interaction energy
can exceed the average kinetic energy per particle thus leading to $\Gamma > 1$. A direct consequence of this is the emergence of solid-like features
such as sustaining low frequency shear modes \cite{PhysRevLett.84.6026} originally predicted in theoretical works \cite{kaw:3552,PhysRevLett.84.6030}
and later realized in laboratory experiments \cite{PhysRevLett.88.175001}. This makes SCYL a good model system to study a range of collective
phenomena in dusty plasma systems.  SCYL's have also been shown to be excellent test-beds for modeling hydrodynamic flows
\cite{PhysRevLett.104.215003,PhysRevLett.106.135001,PhysFluids.24.092002}, self-organization phenomena such as clustering \cite{PhysRevE.64.066402}
and lane formation \cite{PhysRevE.79.041408} in complex plasmas. A significant amount of numerical work has also been done on the viscosity
measurements in SCYL using both equilibrium \cite{saigo:1210,PhysRevLett.88.065002,PhysRevLett.94.185002} and non-equilibrium molecular dynamics
simulations \cite{PhysRevLett.86.1215,PhysRevLett.96.145003} . These studies have confirmed the existence of a viscosity minimum at some crossover
value $\Gamma_c(\kappa)$. The minimum arises due to the competition between ideal and the excess part of the stress tensor. In later works,
visco-elasticity was also quantified using both experiments \cite{PhysRevLett.105.025002,PhysRevE.85.066402} and numerical simulations
\cite{PhysRevE.85.066401} where the crossover frequency for the real and imaginary parts of the complex viscosity was shown to be empirically related
to the inverse Maxwell time. It was shown in Ref \cite{PhysRevE.85.066401} that this crossover frequency develops a maximum at some $\Gamma_c(\kappa)$
implying a minimum in the Maxwell time $\tau_M$ at the same $\Gamma_c(\kappa)$. However, a systematic study of the microscopic origin of shear
relaxation and Maxwell time in these liquids at various ($\Gamma, \kappa$) values is still lacking.  The aim of this letter is to fill this gap and
provide from first principles, an atomistic study that would provide answers to the following two fundamentally important questions: (a) What is the
connection between the microscopic world and the macroscopic shear relaxation in these SCYL? (b) Is $\tau_M$ the duration of dominant elastic response
for these liquids especially at $\Gamma \ll \Gamma_c(\kappa)$? To address the first question, we have calculated the lifetime of local atomic
connectivity $\tau_{LC}$ \cite{PhysRevLett.110.205504} and found that it converges to the relaxation of the excess part of shear stress
auto-correlation $\tau^{ex}_M$ at $\Gamma < \Gamma_c(\kappa)$. Since $\tau_{LC}$ corresponds to the time duration where the topology of nearest
neighbors remains intact, the fact that $\tau^{ex}_M \rightarrow \tau_{LC}$ for $\Gamma < \Gamma_c(\kappa)$ directly indicates that $\tau^{ex}_M$ is the
duration of dominant elastic response at these temperatures. This is different from the results reported in Ref \cite{PhysRevE.85.066401} where
$\tau_{LC}$ was shown to converge to $\tau_M$ below the cross-over temperature for a range of ordinary liquids. It is important to note that none of
those liquids showed a non-monotonous behavior in $\tau_M$ with temperature. We also show that at $\Gamma > \Gamma_c(k)$, $\tau_{LC}$ deviates from
$\tau^{ex}_M$ directly indicating a crossover from a kinetic regime to potential energy dominated regime. Our study further shows that the infinite
frequency shear modulus $G_{\infty}$ does not have a minimum at any temperature as opposed to the viscosity data which has a well known minimum at
$\Gamma = \Gamma_c(\kappa)$.  In the following we provide the details of our numerical work and also explain the procedure used to extract quality
data.

\textit{Numerical simulations:} We have performed molecular dynamics (MD) simulations on a two-dimensional (2D) Yukawa liquid in a canonical ensemble under periodic
boundary conditions. All distances are normalized to Wigner-Seitz radius $a$, energies are normalized to $\frac{Q^2}{4\pi\epsilon_0 a}$ and times are
normalized to $\omega^{-1}_{pd}$.  Here $\omega_{pd}$ is the 2D nominal plasma frequency given by $\omega_{pd} =
\sqrt{\frac{Q^2}{2\pi\epsilon_0ma^3}}$. The simulation box contains 5016 particles at a reduced number density $n = \pi^{-1}$. The dimensions of rectangular
box were chosen to be 125.705291 $\times$ 125.358517 which allows for the formation of a perfect triangular lattice below the freezing transition. The
interaction potential is truncated smoothly to zero along with its first two derivatives by employing a fifth order polynomial as a switching function
in the range ($r_m < r < r_c$) where $r_m$ and $r_c$ are the inner and the outer cutoff respectively. We chose $r_m$ and $r_c$ subject to the
criteria: $\phi(r_m) \approx 2.27 \times 10^{-6}$ and $\phi(r_c) \approx 1.49 \times 10^{-7}$ thus ensuring negligible perturbation to the bare
Yukawa potential. A Nose-Hoover thermostat \cite{PhysRevA.31.1695} with a time constant of $\frac{1}{\sqrt{2}}$ is employed to maintain the
temperature at a desired $\Gamma$. To improve statistics, we have averaged our data over an ensemble of 3200 independent realizations. This was
necessary to reduce the fluctuations present in the long time tail of the stress relaxation function originating from the long range nature of the
interaction potential. We now proceed to explain the computation of various dynamical quantities reported in this paper.

\par
\begin{figure}[h]
  \centering
  \includegraphics[scale=0.23]{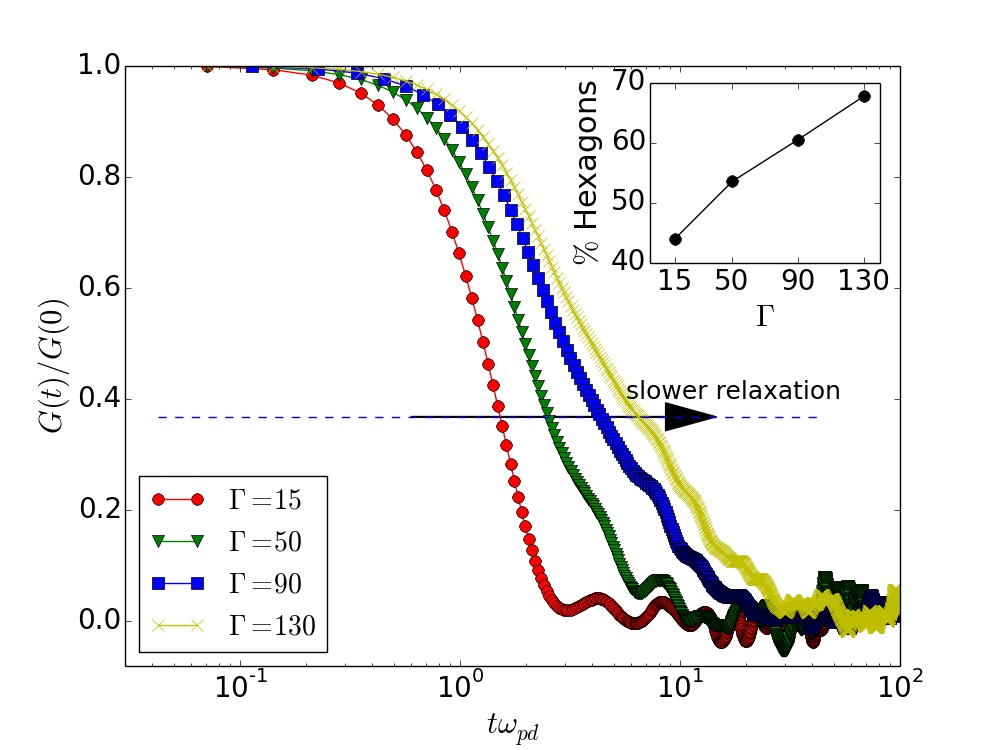}
  \caption{(color online). Normalized stress relaxation function $G(t)$ for a Yukawa liquid at various $\Gamma$. The dashed line shows the location of $1/e$ falling
  time.  It is clear from the figure that the relaxation time increases with $\Gamma$ for $\Gamma > \Gamma_c(\kappa)$. Note that the relaxation time (or
e-folding time) increases with $\Gamma$. [Inset: shows the growing hexagonal order in the liquid with $\Gamma$.]}
  \label{fig1}
\end{figure}

 The stress relaxation function used in our work  is the auto-correlation of the shear stress
tensor and is given as 
\begin{equation}
  G(t) = \frac{1}{A k_B T} \langle \sigma_{xy}(t) \sigma_{xy}(0)\rangle
  \label{eq-sacf}
\end{equation}
with initial value of this auto-correlation giving the infinite frequency shear modulus $G_\infty = G(0)$.
The angular brackets here denote the average over the entire ensemble,
with $A$ being the area and $\sigma_{xy}(t)$ the microscopic stress tensor being defined as
\begin{equation}
  \sigma_{xy}(t) = \sum_{i=1}^N m_i v^x_i(t) v^y_i(t) - \sum_{i=1}^N \sum_{j>i}^N x_{ij}(t) y_{ij}(t)
  \frac{\phi'(r_{ij},t)}{r_{ij}(t)}
  \label{eq-sigma}
\end{equation}
The first term on the right hand side has purely kinetic origins and is the dominating term at high temperatures whereas the second term is the excess
part and has its origins in particle interactions. The second term is named excess as it would be absent in an ideal gas (gas of \textit{``hard
spheres''}) and arises only in a real gas. Figure \ref{fig1} shows the normalized stress relaxation function $G(t)$ of a 2D Yukawa liquid at various
coupling strengths $\Gamma > \Gamma_c(\kappa)$. It is clear that at short times, $G(t)$ has a zero slope meaning that liquid response is dominantly elastic. This is
followed by a region of fast decay where both the elastic and viscous effects are comparable.  At large times, $G(t)$ has become much smaller (within
statistical noise) indicating a regime dominated by viscous response and negligible elastic effects. It is also seen from the figure that the relaxation
time increases with $\Gamma$ implying that elastic response will dominate for longer times as $\Gamma$ increases. This is mainly due to growing
structural order as shown in the inset of Figure 1. Next we turn our attention to the hydrodynamic (or zero frequency) shear viscosity $\eta$ which is an
important dynamical property responsible for viscous dissipation in liquids.

\par All liquids tend to gradually resist the deformation imposed by an external shear stress. A measure of this resistance is the hydrodynamic
shear viscosity $\eta$ (bulk viscosity in case of a compressive external force). For a liquid at thermal equilibrium, 
this shear viscosity can be obtained by the long time integral of the stress relaxation function -a 
procedure well known as the \textit{Green-Kubo formula} \cite{HansenMcDonald}.
\begin{equation}
  \eta = \int^{\infty}_0 G(t) dt
  \label{eq-eta}
\end{equation}
\begin{figure}[h]
  \centering
  \includegraphics[scale=0.23]{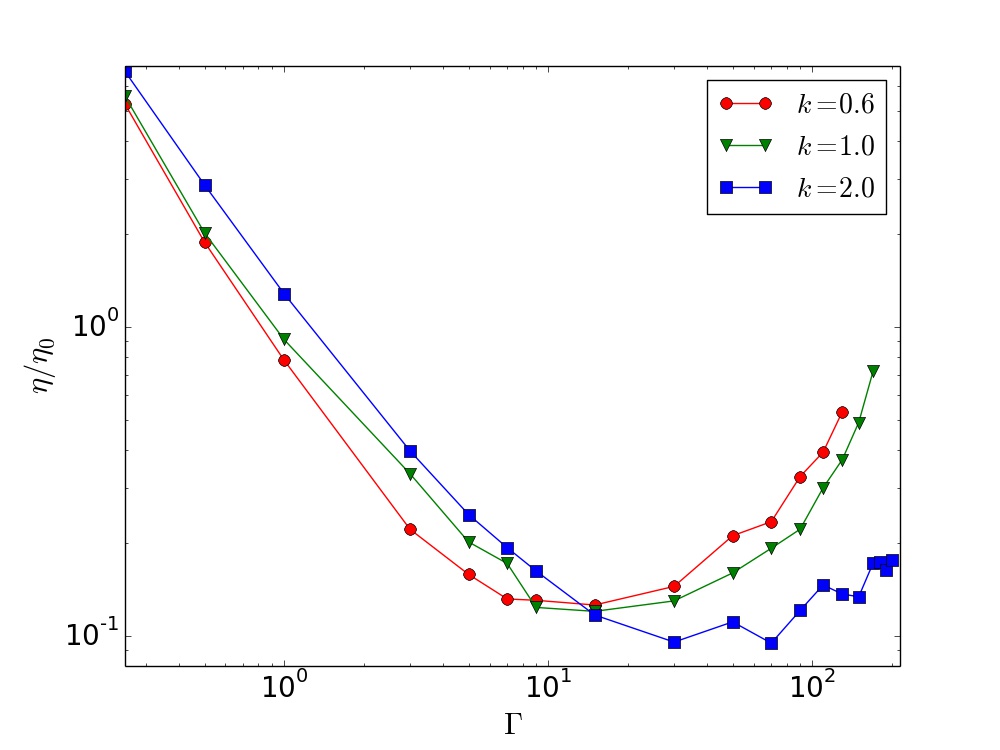}
  \caption{(color online). Hydrodynamic viscosity (normalized to $\eta_0 = a^2 n m \omega_{pd}$) computed from Eq. (\ref{eq-eta}) at various $\kappa$. The minimum
  at $\Gamma = \Gamma_c(\kappa)$ is due to the competition of the ideal part and the excess part of the stress tensor and is well known in the literature
  \cite{saigo:1210,PhysRevLett.94.185002}.}  
  \label{fig2}
\end{figure}
\begin{figure}[h]
  \centering
  \includegraphics[scale=0.23]{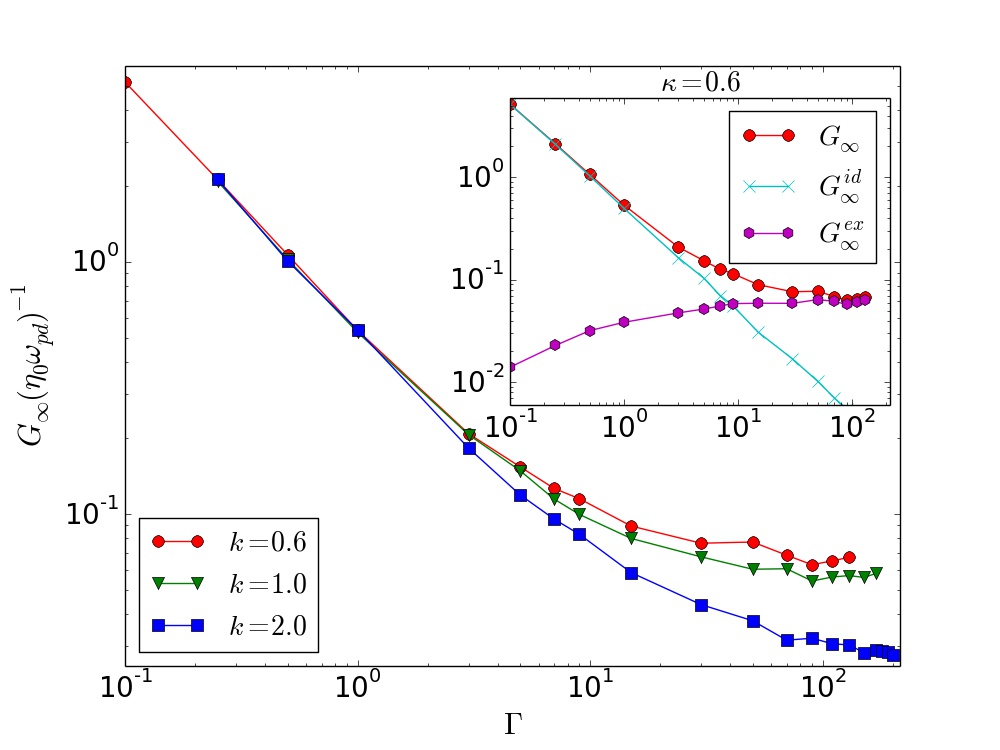}
  \caption{(color online). Infinite frequency shear modulus $G_\infty$ vs. $\Gamma$ at various $\kappa$. Unlike viscosity which has a minimum, $G_\infty$ does not
    have a minimum at any temperature. [Inset: The total, ideal and excess contributions for the case $\kappa = 0.6$ only. It is clear that at low
    $\Gamma$, the entire contribution comes from the ideal term whereas at high $\Gamma$, the excess term is responsible for the entire $G_\infty$.]
  }
  \label{fig3}
\end{figure}

\par In calculating $\eta$ from Eq. (\ref{eq-eta}), the upper limit of the integral is set to the first zero crossing time of the ensemble averaged shear
relaxation function $G(t)$. In Figure \ref{fig2}, we have shown our data on the shear viscosity measurement at three different values of $\kappa$. Our
results for $\eta$ are both qualitatively and quantitatively similar to the existing report on 2D equilibrium simulations
\cite{PhysRevLett.94.185002}. Next we show our data on the $G_\infty$ calculation in Figure \ref{fig3} at various $\kappa$. It is interesting to see
that unlike shear viscosity, $G_\infty$ does not have a minimum at any temperature. The contributions coming from both the ideal and excess part of
$G_\infty$ are shown in the inset of Figure \ref{fig3}. At high temperatures (low $\Gamma$), kinetic effects are dominant and the ideal part
$G^{id}_\infty$ is the dominating term and becomes a major contributor to the overall $G_\infty$. At low temperatures (high $\Gamma$) the excess term
$G^{ex}_\infty$ begins to dominate and the ideal term becomes very small. It is interesting to note that the excess part $G^{ex}_\infty$ saturates as
$\Gamma$ increases. In the following, we will show the effect of quantities addressed so far in the determining the Maxwell shear stress relaxation
time $\tau_M$.

\begin{figure}[h]
  \centering
  \includegraphics[scale=0.23]{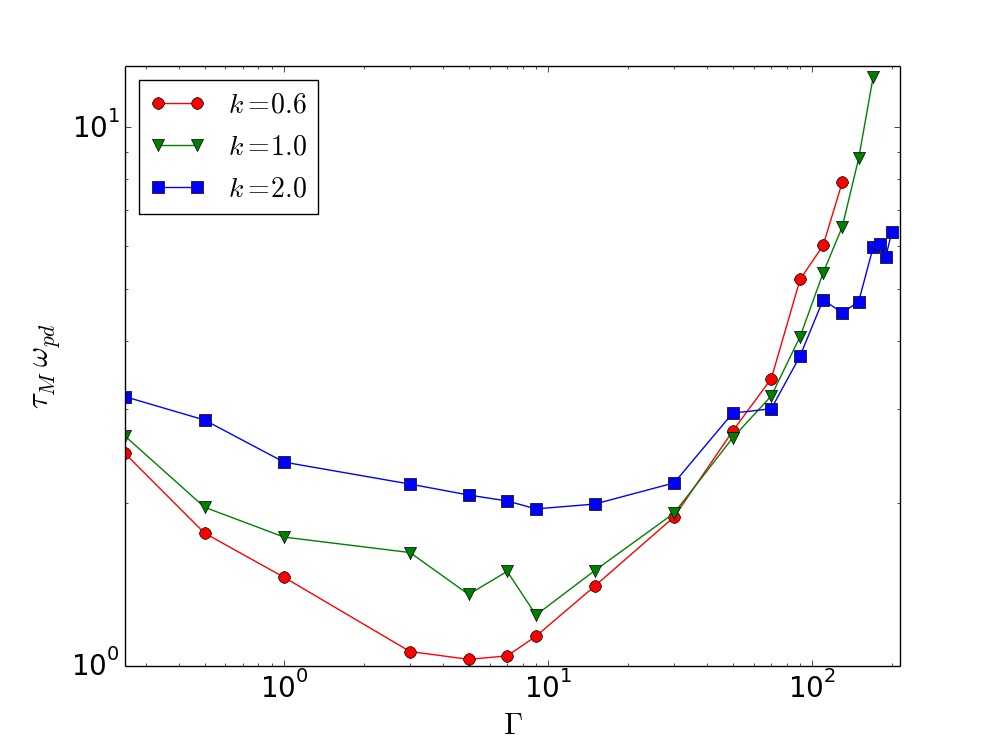}
  \caption{(color online). Maxwell relaxation time $\tau_M$ as calculated from Eq. (\ref{eq-tauM}). The data shows a clear minimum around a crossover
  $\Gamma_c(\kappa)$ as previously shown in Ref \cite{PhysRevE.85.066401}. We explain this non-monotonous behavior by measuring independently the
  relaxation times of the ideal and excess parts of the shear stress auto-correlation function [See text and also Fig. \ref{fig5} for details].}  
  \label{fig4}
\end{figure}

\par It is generally accepted that all liquids are visco-elastic in nature meaning that their mechanical response to an external force will be both viscous
and elastic at the same time. This can be explained through the concept of the Maxwell relaxation time scale $\tau_M$ such that at times $t <<
\tau_M$, the response of the liquid will be dominantly elastic (\textit{reversible}) and at $t >> \tau_M$ the response will be dominantly viscous
(\textit{irreversible}). At intermediate time scales, both elastic and viscous features will be comparable.  The ratio of the hydrodynamic shear
viscosity $\eta$ to the infinite frequency shear modulus $G_\infty$ defines an average shear relaxation time or the Maxwell time
\cite{MountainZwanzig-66},
\begin{equation}
  \tau_M = \frac{\eta}{G_\infty} = \frac{\int^{\infty}_0 G(t) dt}{G_\infty}
  \label{eq-tauM}
\end{equation}
\begin{figure}[h]
  \centering
  \includegraphics[scale=0.23]{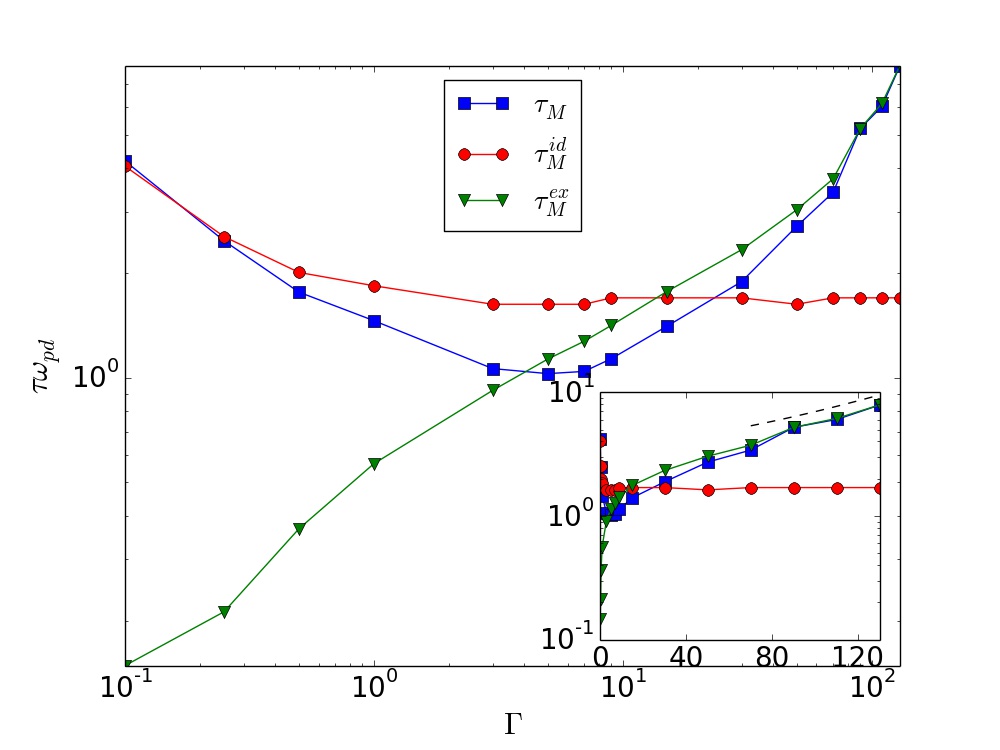}
  \caption{(color online). The relaxation times of the stress relaxation function $G(t)$ along with its ideal $G^{id}(t)$ and excess parts $G^{ex}(t)$ for the case of
  $\kappa = 0.6$. It is interesting to note that while the excess term $\tau^{ex}_M$ continues to rise all the way to the freezing transition,
  the ideal part saturates to a constant value around $\Gamma_c(\kappa)$ indicating the onset of potential energy influenced regime. 
  [Inset: Data plotted on a log-linear axis. A dashed exponential is drawn to aid the eye of the reader.]}
  \label{fig5}
\end{figure}
\begin{figure}[h]
    \centering
    \includegraphics[scale=0.20]{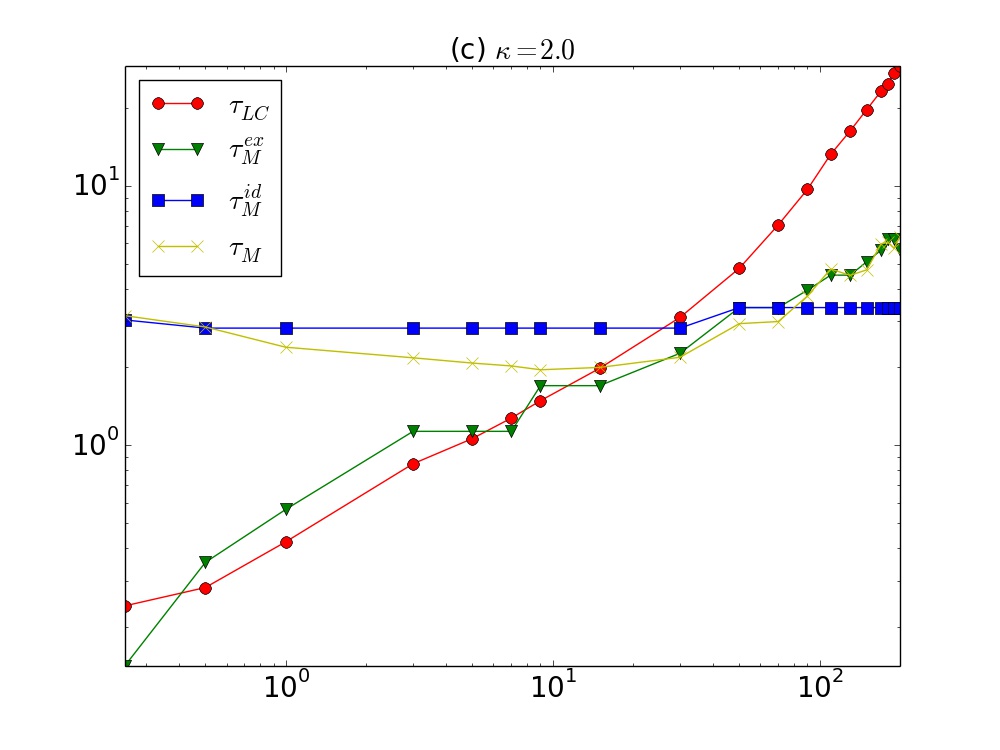}
    \includegraphics[scale=0.20]{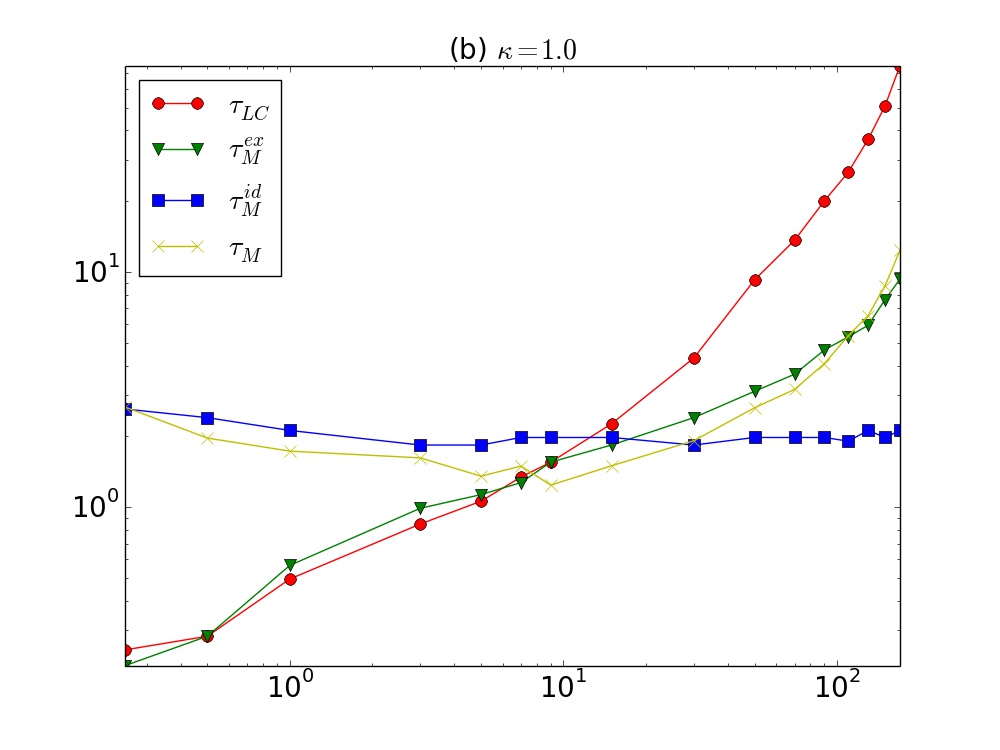}
    \includegraphics[scale=0.20]{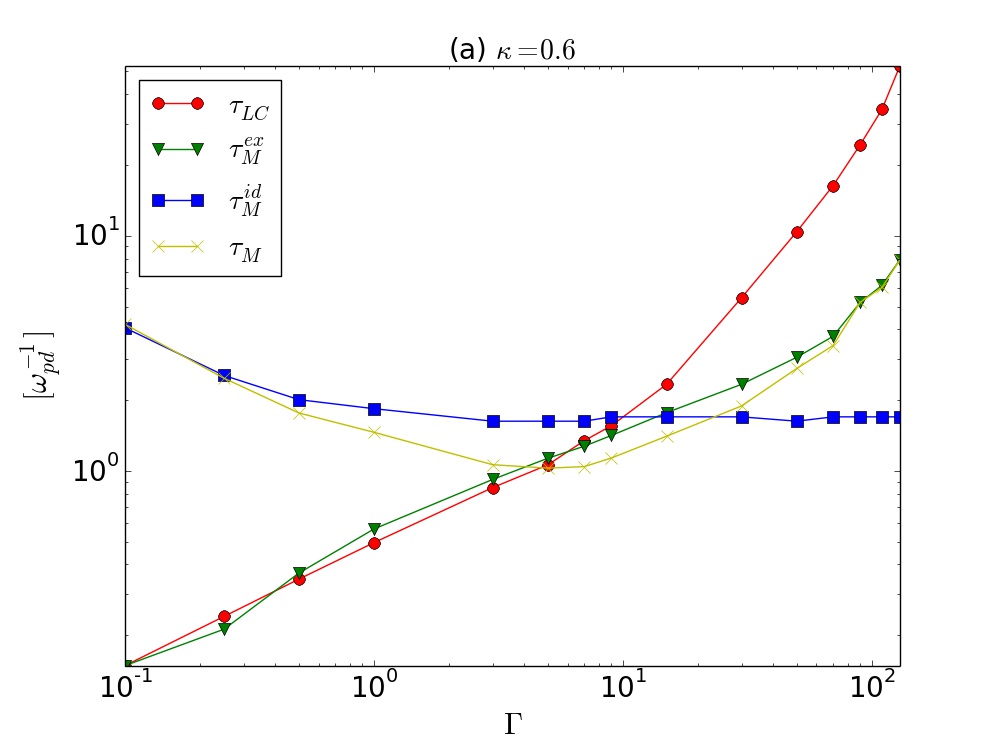}
    \caption{(color online). Comparison of $\tau_{LC}$ with time scales $\tau_M$, $\tau^{ex}_M$ and $\tau^{id}_M$ for the cases (a) $\kappa =
    0.6$, (b) $\kappa = 1.0$, (c) $\kappa = 2.0$. At $\Gamma < \Gamma_c(\kappa)$, $\tau^{ex}_M$ becomes equal to the lifetime of atomic connectivity $
    \tau_{LC}$. At $\Gamma > \Gamma_c(\kappa)$ these two time scales deviate as the liquid enters a landscape dominated regime where the 
    interaction between local networks becomes significant and leads to cancellation of long range elastic fields
    \cite{PhysRevLett.110.205504}}. 
 \label{fig6}
\end{figure}

The normalized integral given above provides the duration of the auto-correlation and is hence a good measure of the time duration of $G(t)$. Figure
\ref{fig4} shows a plot of $\tau_M$ vs. coupling strength $\Gamma$ for three values of $\kappa$. Our data shows the minimum in $\tau_M$ at $\Gamma =
\Gamma_c(\kappa)$ similar to the results reported in Ref \cite{PhysRevE.85.066401}. This means that if one considers $\tau_M$ (the relaxation of the
total stress auto-correlation) as a measure of elastic response then it would imply that elasticity will persist for longer times as the weakly
coupled limit or $\Gamma \rightarrow 0$ is approached. We thus propose that $\tau_M$ would be an incorrect measure of elasticity in these liquids as
$\Gamma \rightarrow 0$.  To seek for an appropriate measure of elasticity we turn our attention to the components of the total stress auto-correlation
$G(t)$ namely the ideal part $G^{id}(t)$ and the excess part $G^{ex}(t)$. There is a cross term as well but its value is negligible at all values of
$(\kappa, \Gamma)$. In figure \ref{fig5} we show the relaxation times of the total total stress auto-correlation and its two components over the range
of $\Gamma$ that spans from the very weakly coupled fluid limit $\Gamma = 0.25$ to the highest $\Gamma $ just below freezing transition.  As the
liquid nears freezing ($\Gamma \rightarrow \Gamma_m$), $\tau^{ex}_M \rightarrow \tau_M$ and the shear relaxation is dominantly due to the particle
interactions. We find that the relaxation time of the $G^{ex}(t)$ denoted here by $\tau^{ex}_M$ serves as a good measure of elasticity as it goes to
zero monotonically as $\Gamma \rightarrow 0$. On the high $\Gamma$ side, it rises exponentially with $\Gamma$ (see inset: Figure \ref{fig5}) as the
liquid approaches the freezing transition. On the other hand, as $\Gamma \rightarrow 0$, relaxation time of $G^{id}(t)$ denoted here as $\tau^{id}_M$
becomes dominant and is entirely responsible for the origin of $\tau_M$. This means that in the weakly coupled fluid limit, stress relaxation is
purely kinetic and has very little contribution coming from particle interactions. It is very important to note here that our assertion on
$\tau^{ex}_M$ being the right indicator for elasticity is not just based on the right asymptotic behavior with $\Gamma$ but also due to its
convergence to the lifetime of local atomic connectivity $\tau_{LC}$ as explained in the following section.

\par In solids, phonons are weakly scattered (long lived) and hence they are the microscopic origin for vibrations. In liquids however, they
exhibit highly marginalized behavior as they are strongly scattered due to lack of any underlying long range structural order. As a result phonons
cannot be used to explain the microscopic origins of viscosity and the infinite frequency shear modulus. One is thus led to the following
fundamentally important question: What is the microscopic origin for shear relaxation in SCYL? To answer this question,
we measure the lifetime of the local atomic connectivity $\tau_{LC}$ \cite{PhysRevLett.110.205504} in SCYL.  To calculate $\tau_{LC}$ one first assigns a set
of bonds between a central atom and its nearest neighbors at some reference time $t_0$. The nearest neighbors here correspond to atoms situated at a
distance lesser than the location of the first minima in the radial distribution function. Once a reference state is assigned we track the change in
local atomic connectivity i.e. the central atoms \textit{losing} or \textit{gaining} some neighbors. As the simulation proceeds the average
coordination number will reduce and $\tau_{LC}$ is extracted as the time duration (elapsed since $t_0$) in which the average coordination falls by 1.
To improve statistics, we averaged our data over several independent realizations. In this sense, $\tau_{LC}$ is the average time during which any atom
will lose one neighbor. It was shown in Ref \cite{PhysRevLett.110.205504} that above a certain crossover temperature,  $\tau_{LC}$ 
converges to the Maxwell time $\tau_M$ and is thus the microscopic origin of shear relaxation in a range of liquids. It should be noted
that none of the liquids used in that work showed a non-monotonic behavior in $\tau_M$. In Figure \ref{fig6} we show our data on $\tau_{LC}$
calculated at various values ($\Gamma, \kappa$) and make the following interesting observation. At $\Gamma < \Gamma_c(\kappa)$, $\tau_{LC} \rightarrow
\tau^{ex}_M$ making $\tau_{LC}$ the microscopic origin of excess part of stress at these temperatures. Since $\tau_{LC}$ corresponds to the time
duration where the topology of nearest neighbors remains intact it is also the duration of the dominant elastic or ``solid like'' response. These
observations directly imply that the relevant shear relaxation time scale which quantifies the elastic response in these liquids is $\tau^{ex}_M$
which becomes equal to $\tau_{LC}$ at $\Gamma < \Gamma_c(\kappa)$ and approaches the value of $\tau_M$ at  $\Gamma > \Gamma_c(\kappa)$. The reader
should also note that at $\Gamma > \Gamma_c(k)$, $\tau_{LC}$ deviates from 
$\tau^{ex}_M$ directly indicating a crossover from a kinetic regime to potential energy dominated regime. This is a regime where interaction between
local networks becomes significant and leads to cancellation of long range elastic fields \cite{PhysRevLett.110.205504}.

\par \textit{Summary:} We have calculated the infinite frequency shear modulus $G_\infty$, Maxwell relaxation times ($\tau_M, \tau^{id}_M \text{and}
\tau^{ex}_M$) and the lifetime of the local atomic connectivity $\tau_{LC}$ of a 2D SCYL using accurate MD simulations.  The simulations were done
under a canonical ensemble and quality data is obtained by averaging the runs over 3200 independent realizations. Our major finding is the microscopic
origin of shear relaxation in SCYL. We find that the lifetime of topology of nearest neighbors $\tau_{LC}$ is responsible for the relaxation of the excess part of
the shear stress relaxation $\tau^{ex}_M$. We also provide a solution to the riddle concerning the minimum in $\tau_M$ which renders it unsuitable to quantify the
elastic properties of a SCYL approaching the weakly coupled fluid limit $\Gamma \rightarrow 0$.  We resolve this issue by proposing that the  correct
object that quantifies elastic response of such strongly coupled liquids is the relaxation time of the excess part of the shear stress
auto-correlation $\tau^{ex}_M$ which goes to zero as $\Gamma \rightarrow 0$ and becomes exponential as the liquid nears the freezing transition. Our
assertion is backed by careful measurements of the lifetime of the local atomic connectivity $\tau_{LC}$ which clearly shows that $\tau^{ex}_M
\rightarrow \tau_{LC}$ below the crossover $\Gamma_c(\kappa)$ directly indicating that $\tau^{ex}_M$ is the duration of dominant elastic response. The
relaxation of the ideal part of shear stress auto-correlation $\tau_{id}$ saturates with $\Gamma > \Gamma_c(\kappa)$ implying  a crossover
from the kinetic regime to the potential energy regime. We also find that the infinite frequency shear modulus
$G_{\infty}$ does not have a minimum at any temperature as opposed to the viscosity data which has a well known minimum around $\Gamma_c(\kappa)$.

\par We believe the present work is fundamentally important as it provides the microscopic origin of shear relaxation time scales in a SCYL.  The
results presented here may also apply to other models of visco-elastic liquids exhibiting a non-monotonic behavior of Maxwell time with
temperature.  It will be very interesting to extend the present study to the binary super-cooled Yukawa liquids.
\begin{acknowledgments}
    Discussions with P.K. Kaw and R. Ganesh are gratefully acknowledged. This work was
      supported under the INSPIRE faculty program, Department of Science and Technology, Ministry of
      Science and Technology, Government of India.
\end{acknowledgments}

\bibliography{ashwin}

\end{document}